\documentclass[apj]{emulateapj}
\usepackage{color}

\shorttitle{From Dust to Planetesimal: the Snowball Phase ?}
\shortauthors{Xie et al.}

\begin{document}

 \title{From Dust to Planetesimal: the Snowball Phase ?}

\author{Ji-Wei Xie$^{1,2}$, Matthew J. Payne$^2$, Philippe Th{\'e}bault$^3$, Ji-Lin Zhou$^1$,  Jian Ge$^2$, }
\affil{$^1$Department of Astronomy, Nanjing University,
Nanjing, Jiangsu, 210093, China} 
\affil{$^2$Department of Astronomy, University of Florida, Gainesville, FL, 32611-2055, USA}
\affil{$^3$Observatoire de Paris, Section de Meudon, F-92195 Meudon Principal Cedex, France}
\email{xiejiwei@gmail.com}

\begin{abstract}
The standard model of planet formation considers an initial phase in which planetesimals form from a dust disk, followed by a phase of mutual planetesimal-planetesimal collisions, leading eventually to the formation of planetary embryos. However, there is a potential transition phase (which we call the ``snowball phase"), between the formation of the first planetesimals and the onset of mutual collisions amongst them, which has often been either ignored or underestimated in previous studies.  In this snowball phase, isolated planetesimals move on Keplerian orbits and grow solely via the direct accretion of sub-cm sized dust entrained with the gas in the protoplanetary disk. Using a simplified model in which planetesimals are progressively produced from the dust, we consider the expected sizes to which the planetesimals can grow before mutual collisions commence and derive the dependence of this size on a number of critical parameters, including the degree of disk turbulence, the planetesimal size at birth and the rate of planetesimal creation. For systems in which turbulence is weak and the planetesimals are created at a low rate and with relatively small birth size, we show that the snowball growth phase can be very important, allowing planetesimals to grow by a factor of $10^6$ in mass before mutual collisions take over. In such cases, the snowball growth phase can be the dominant mode to transfer mass from the dust to planetesimals. Moreover, such growth can take place within the typical lifetime of a protoplanetary gas disk. 
A noteworthy result is that, for a wide range of physically reasonable parameters, mutual collisions between planetesimals become significant when they reach sizes $\sim 100$ km, irrespective of their birth size. This could provide an alternative explanation for the turnover point in the size distribution of the present day asteroid belt.
For the specific case of close binaries such as $\alpha$ Centauri, the role of snowball growth could be even more important. Indeed, it provides a safe way for bodies to grow through the problematic $\sim $1 to 50\,km size range for which the perturbed environment of the binary can prevent mutual accretion of planetesimals.
From a more general perspective, these preliminary results suggest that an efficient snowball growth phase provides a large amount of ``room at the bottom" for theories of planet formation.
\end{abstract}

%

\section{Introduction}\label{Intro}
The standard core-accretion model generally treats planet formation as occurring in two stages \citep{Lis 93, Cha 04, Arm 10}: In the first stage, mountain-sized planetesimals form from small dust grains embedded in a gas-rich protoplanetary disk \citep{Wei 97, BW 08, You 08, CY 09}, and then in the second stage these planetesimals go on to accrete one another to form planetary embryos \citep{KI 96, KI 98, Wei 97} which eventually go on to merge into full planets \citep{CW 98, LA 03, Kok 06}. 
The first stage - planetesimal formation - is of crucial importance, as it sets the initial conditions upon which subsequent stages of evolution depend. However, the details of planetesimal formation are still poorly understood and remain somewhat controversial. 

One model for the formation of planetesimals envisages them growing via mutual sticking collisions \citep{WC 93, Wei 97}. Laboratory experiments show that $\mu$m 
sized dust grains can stick together efficiently through various surface forces, including the van der Waals and electrostatic forces, causing them to form mm to cm-sized aggregates \citep{BW 08}. Beyond this mm to cm range, particles become less sticky while their gravitational interactions remain weak, leading to collisional disruption rather than growth \citep{Dom 07}. Furthermore,  cm to m sized objects begin to decouple from the gas, so that they experience collisions with higher relative velocities and also begin to experience significant aerodynamic drag which causes them to quickly spiral into the protostar on a timescale of a few hundred orbital periods \citep{Wei 77} -- the well-known ``meter-barrier".

An alternative scenario that has been suggested to try and circumvent this meter-barrier, envisages that the km-sized planetesimals form directly via gravitational instabilities in a dense particle sub-disk near the mid-plane of the protoplanetary disk \citep{Saf 69, GW 73, YS 02}. However, it has been pointed out that even low levels of turbulence would prevent solids from settling down to a sufficiently dense mid-plane layer \citep{Wei 80, CW 06}. 

Recently breakthroughs have been made in both scenarios.
For the collisional scenario, \citet{TW 09} found in experiments that high-velocity (tens of $\rm m.s^{-1}$) collisions between small dust particles ($\rm<cm$) and a pre-planetesimal body ($\rm>dm$) can lead to efficient growth, suggesting that some ``lucky" pre-planetesimals may escape from erosive collisions and go on to form km-sized planetesimals \citep{Joh 08}. 

For the instability scenario,  it is found that turbulence can act to concentrate dust, thus providing a high density, low velocity dispersion environment in which planetesimals may form via various models of instability \citep{Joh 07, Cuz 08}. The planetesimals formed in these new turbulent models can be 1-2 orders of magnitude larger than those formed in traditional models. 

Planetesimal formation is thus far from being completely understood and current models are a work-in-progress. However, a point worth mentioning is that none of these different scenarios ensure that all planetesimals appear at the same time at a given location. In fact, any model which requires non-global density enhancements to form planetesimals, e.g. \citet{Joh 07} and \citet{Cuz 08}, implicitly assumes that certain regions of the disk will form planetesimals first, while the rest of the disk remains ``dusty". 
Furthermore, two recent studies by \citet{Cha 10} and \citet {Cuz 10} argue that, for the turbulent concentration scenario, planet formation efficiency (set by the probability for turbulently-formed clumps of mm-sized grains to collapse into planetesimals) could be relatively low, depending on the values of several crucial parameters such as the dust-to-gas density ratio, disk viscosity, density and profile. This means that, at a given location in the disk, there could be a wide time gap between the moment when the first planetesimals form and the moment when most of the available solid mass has been converted into planetesimals.

There could thus be a long transition period during which individual planetesimals are embedded in a disk where most of the mass of solids is still in small grains. This issue has tended to be ignored in studies investigating the next stage of planet formation, i.e., planetesimal accretion. Most of these studies consider a system where all planetesimals are present at time $t_0$, possibly with a distribution of initial sizes, which then grow by pairwise accretion \citep{Wei 97, KI 96, KI 98, Bar 09}. Nevertheless, there are several noteworthy exceptions. \citet{WI 00} considered that planetesimals progressively appear over a $10^{5}$years timescale, but only considered mutual planetesimal accretion as a possible growth mode, implicitly neglecting the contribution of dust. \citet{Mor 09}, while attempting to fit the known asteroidal size-distribution, did consider in one of their simulations the possible accretion of dust by planetesimals which formed early, but did so only for one specific case: large, 100\,km sized seed-planetesimals produced over 2 Myr. In addition, \citet{LR 05} attempted to model accretion of dust onto planetesimals as well as planetesimal-planetesimal collisions but they started with very large planetesimals. The most promising study has been recently performed by \citet{PL 10}, who envisage dust accretion onto planetesimals more explicitly, but restricted to the specific case of a close-in binary, and using only a simplified 2-dimensional model. 

In this study, we plan to take these pioneering studies a step further and investigate in detail planetesimal growth during the transition period when isolated planetesimals and a primordial dust-disk coexist. We outline a new growth mode, which dominates during this early phase, during which planetesimals experience negligible gravitational or collisional interactions with one another, and grow mainly (or solely) via the accretion of dust or ice that they sweep up$-$in the manner of a rolling snowball. In this paper, we refer to this dust-fueled planetesimal growth phase as the ``snowball" growth phase. 

The paper is organized as follows. The snowball growth rate (growth only by dust accretion) is derived, using a semi-analytical approach, in $\S\,$2. Next, in $\S\,$3, we show how long the snowball phase could last and to what radial size and mass fraction the planetesimals can grow via this snowball growth mode. Then, in $\S\,$4, the implications for planet formation in both the Solar System and in close binary systems are discussed. Finally, we conclude in $\S\,$5.  Additionally, all variables defined in this paper are listed  alphabetically in table \ref{tbl-1}.

 
\section{Model} \label{model}

\subsection{Planetesimal Formation Rate} \label{pfrate}
We adopt a simplified analytical model which contains only two components: dust and planetesimals. The dust in our model is assumed to be in the sub-mm to mm size range , as (1) this is the typical size of chondrules \citep{Sco 07}, and (2) particles of this approximate size settle down to the mid-plane of the disc on a relative short timescale of $\sim10^3$ yr, and can survive the inward drift for as long as $\sim10^6$ yr \citep{You 08}. 

{\it Individual} planetesimals are assumed to be quickly produced from dust, on a timescale $t_{f.ind}$, whether through collisions \citep{TW 09} or instabilities \citep{Joh 07, Cuz 08}.  
A reasonable reference value for $t_{f.ind}$ might be $\sim10^4$ yr \citep{Lis 93}. However, we do \emph{not} restrict our study to $t_{f.ind}=10^4$ yr. In fact, as we show in section 3, the precise value of $t_{f.ind}$ is unimportant to the final results as long as it is not too long (i.e.  $t_{f.ind}<t_{snow}$, see $\S\,$3). 

We make the simplifying assumption that, once formed from the dust, all planetesimals have the same size $R_{p0}$, which we take as the initial condition for newly-formed planetesimals in our model. To account for the dispersion in times at which planetesimals are created in the system, we introduce an efficiency factor $\epsilon_p$ ($0<\epsilon_p\le1$) for planetesimal formation at a given location in the protoplanetary disk. We then  consider a given radial distance in the disk and define the local planetesimal formation rate as
\begin{equation}
\frac{dN}{dt}=\epsilon_p\times {\Sigma_{d}\over M_{p0}}\left( {1\over t_{f.ind}} \right)={\Sigma_{d}\over M_{p0}}\left( {1\over t_{f}} \right)
\label{rate},
\end{equation}
where $N$ and $M_{p0}$ are the surface number density and initial {\it individual} mass of the planetesimals respectively, and $\Sigma_d$ is the dust surface density. Further, following \citet{Cha 10}, $t_f$ is the characteristic planetesimal formation timescale defined as
\begin{equation}
t_f = \Sigma_{d} \left(M_{p0} \frac{dN}{dt} \right)^{-1}=\frac{t_{f.ind}}{\epsilon_p}.
\label{tform}
\end{equation}
 The surface number density of planetesimals in the system thus increases linearly according to the relation
\begin{equation}
N=\epsilon_p\times {\Sigma_{d}\over M_{p0}}\left( {t\over t_{f.ind}} \right)={\Sigma_{d}\over M_{p0}}\left( {t\over t_{f}} \right),
\label{rate2}
\end{equation}
 assuming that $N=0$ at $t=0$. 
 
 Note that $t_f$ does \emph{not} represent the time it takes for an individual planetesimal to form (that is $t_{f.ind}$) but rather the time it takes for most of the dust to be converted into planetesimals, as parameterized by the factor $\epsilon_p$. If $\epsilon_p=1$, i.e., $t_f=t_{f.ind}$, this implies that all the planetesimals are created at the same instant. This should be treated as an idealized or limiting case, and in reality, we expect $\epsilon_p<1$, i.e., $t_f>t_{f.ind}$. \citet{Cha 10} finds that $t_f$ can easily vary over 10 orders of magnitude, from $10^{4}$ to $10^{14}$ yr, depending on local conditions in the protoplanetary disk.

Our model, assuming a constant rate of planetesimal creation as well as a single initial planetesimal size, is very simplified. However, it is accurate enough for the order-of-magnitude approach adopted in this present study and allows us to identify and quantify the snowball phase in a convenient manner.

\subsection{Growth Rate via Dust Sweeping}
When a planetesimal is sweeping through a dust disk, the mass growth rate through dust accretion is  
\begin{equation}
\dot{M}_p=E_d \pi R_p^2{\Sigma_d\over2H_d}v_{rel} ,
\label{sweep}
\end{equation}
where $M_p$, and $R_p$ are the mass and radii of the planetesimal, $H_d$ is the scale height of the dust-disk, and $v_{rel}$ is the relative velocity between planetesimal and dust.
$E_d$ is a dimensionless coefficient accounting for the efficiency of dust accretion onto planetesimals. According to \citet{TW 09}, accretion can be very efficient as long as the dust size is smaller than mm. Hence, throughout this paper, we take a medium value $E_d=0.5$. 
From Eqn.\ref{sweep}, we can also derive the growth rate in the planetesimal radius as
\begin{equation}
\dot{R}_p={E_d\over8}\left({\Sigma_d\over H_g\rho_*}\right)\left({H_g\over H_d}\right)v_{rel},
\label{grate}
\end{equation}
where $\rho_*$ is the planetesimal density and $H_g$ is the scale height of the gas-disk. Throughout this paper, we adopt $\rho_*=3\rm g.cm^{-3}$ and use a protoplanetary disk scaled by the Minimum Mass Solar Nebula (MMSN, \citet{Hay 81}), which gives $\Sigma_d\sim10f_d f_{ice} ({a/\rm AU})^{-3/2} \rm g.cm^{-2}$, and $H_g\sim0.05({a/ \rm AU})^{5/4}\rm AU$, where $f_d$ is the scaling factor of the disk mass relative to the MMSN, and $f_{ice}$ accounts for the enhancement of solid density beyond the ice line, $a_{ice}$. We take $f_{ice}=4.2$ if $a>a_{ice}$, and $f_{ice}=1$ otherwise \citep{IL 04}.
Note that in Eqn.\ref{sweep} and Eqn.\ref{grate}, we do not consider the gravitational focusing effect, since dust is well coupled with the gas while the focusing effect is only well applied to 2-body dynamics in which only the gravity of the two objects themselves matters.  
As a consequence,  $v_{rel}$ is basically the local differential velocity between a large planetesimal and a gas streamline. Assuming an axisymmetric and pressure-supported gas disk, this leads to
\begin{equation}
v_{rel}\sim ({H_g/a})^2v_k\sim75 \rm m.s^{-1},
\label{vrel}
\end{equation}
where $v_k$ is the local Keplerian velocity at $a$ AU.
Substitutinging this into Eqn.\ref{grate}, we get the growth rate in the planetesimals' radii due to the direct accretion of dust, 
\begin{equation}
\dot{R}_p\sim7\times10^{-7}f_d f_{ice}\left({H_g\over H_d}\right)\left({a\over AU}\right)^{-11/4} \ \ \ \rm km.yr^{-1}.
\label{grate2}
\end{equation}
Taking $f_d\times f_{ice}\sim 1-10$ and $H_g/H_d\sim1-1000$, we get $\dot{R}_p\sim10^{-6}-10^{-3} \rm km.yr^{-1}$ at 1 AU.
 
It is worth noting  from Eqn.\ref{grate2} that the planetesimal radii follow a linear growth rate.
Since the planetesimal formation rate is also assumed linear as shown in Eqn.\ref{rate2}, then we can derive the mass-weighted average  planetesimal radius as (see Appendix for detail)
\begin{equation}
\langle R_p\rangle\sim R_{p0} +\left({1\over4}\right)^{1/3}\dot{R}_pt,
\label{rwei}
\end{equation}
as well as the Planetesimal Mass Fraction (PMF) with respect to the total solid mass,
\begin{equation}
PMF=\epsilon_p\left({\langle R_p\rangle\over R_{p0}}\right)^3\left({t\over t_{f.ind}}\right)=\left({\langle R_p\rangle\over R_{p0}}\right)^3\left({t\over t_{f}}\right).
\label{pmf}
\end{equation}
$\langle R_p\rangle$ and $PMF$ are two key statistics which trace the typical size and total mass of planetesimals in the protoplanetary disk. Note that, for the growth rate we derive in Eqn.\ref{grate2}, a constant dust surface density is implicitly assumed. In reality, dust would be consumed both through the production of new planetesimals and through the growth of planetesimals which formed earlier. To account for this depletion, we use $PMF$ as an alarm or flag: if $PMF>0.5$, we then understand that a large fraction of dust has been consumed and we turn off planetesimal growth via dust accretion.

\subsection{Growth via Mutual (Planetesimal-Planetesimal) Accretion}

As planetesimals begin to populate the disk, they can begin having mutual encounters. The timescale for the onset of mutual collisions decreases as the number density increases. Assuming equipartition between in-plane and out-of-plane motions, we get
\begin{eqnarray}
t_{col} &\sim& 1/\left(n \Delta V\pi \langle R_p\rangle^2\right) \nonumber \\
&\sim& {2\over 3\pi}\left({R_{p0}^3\rho_*\over \langle R_p\rangle^2\Sigma_d}\right)\left({t\over t_{f}}\right)^{-1}\left({a\over \rm AU}\right)^3 \ \ \rm yr,
\label{tcol}
\end{eqnarray}
where $n\sim N/(2ai_p)$ and $\Delta V\sim2i_pv_k$ are the volume number density and average relative velocity of the planetesimals, respectively, and $i_p$ is their average orbital inclination.

%
\section{Snowball Growth}

We define snowball growth as beginning with the formation of the first planetesimal and ending at the point when mutual planetesimal collisions take over as the dominant growth mode, i.e., when $t=t_{col}$. If the snowball growth is efficient, i.e., $\langle R_p\rangle \gg R_{p0}$, then the duration of the snowball phase $t_{snow}$ can be approximately derived as 
$$
t_{snow}\sim 10^4 \left(f_d f_{ice}\right)^{-1/4}\left({t_{f}\over \rm 10^4 yr}\right)^{1/4}\left({R_{p0}\over \rm km}\right)^{3/4}
$$
\begin{equation}
\left({\dot{R}_p\over \rm 10^{-4}km.yr^{-1}}\right)^{-1/2}\left({a\over \rm AU}\right)^{3/4} \ \ \rm yr.
\label{tsnow}
\end{equation}  
Given $t_{snow}$ and using Eqn.\ref{rwei} and Eqn.\ref{pmf}, the mass-weighted size ($R_{snow}$) and the mass fraction of planetesimals ($PMF_{snow}$) at the end of the snowball growth phase can be estimated as,
\begin{eqnarray}
R_{snow}&\sim & R_{p0}+0.7\left(f_d f_{ice}\right)^{-1/4}\left({t_{f}\over \rm 10^4 yr}\right)^{1/4}\left({R_{p0}\over \rm km}\right)^{3/4} \nonumber \\
&& \times \left({\dot{R}_p\over \rm 10^{-4}km.yr^{-1}}\right)^{1/2}\left({a\over \rm AU}\right)^{3/4} \label{rsnow} \ \rm km, \\  \textrm{and} \nonumber && \\
PMF_{snow} & \sim & 2.4\times10^{-3}\left({H_g\over H_d}\right)\left({a\over \rm AU}\right)^{1/4} \label{pmfsnow}
\end{eqnarray} 
We wish to emphasise that:
\begin{enumerate}
\item Eqn.\ref{tsnow}, Eqn.\ref{rsnow} and Eqn.\ref{pmfsnow} are approximate solutions which are valid only if $\langle R_p\rangle \gg R_{p0}$. When this approximation is invalid ($\langle R_p\rangle \sim R_{p0}$), then $t_{snow}$, $R_{snow}$ and $PMF_{snow}$ cannot be expressed analytically and numerical solutions are required (see Fig.\ref{figtrsnow} in \S\,3.3).
\item Eqn.\ref{tsnow} and Eqn.\ref{rsnow} depend on $t_f$ (not $t_{f.ind}$). The coefficient of $10^4$ yr appearing in Eqn.\ref{tsnow} and Eqn.\ref{rsnow} is \emph{not} related to the value $t_{f.ind}\sim10^4$ yr mentioned in section 2.1. In fact, $t_{f.ind}$ does not affect the results (Eqn.\ref{tsnow}, Eqn.\ref{rsnow} and Eqn.\ref{pmfsnow}) provided that $t_{f.ind}<t_{snow}$.
\item $PMF_{snow}$ is the total $PMF$ at the end of snowball phase, i.e. $t=t_{snow}$, it is \emph{not} just the $PMF$ contributed \emph{purely} by the snowball \emph{growth} phase. (See also in \S\, 3.1 and \S\, 4.1.1 for details) 
\item Since $\dot{R}_p$ is proportional to $f_d$ and $f_{ice}$, $R_{snow}$ actually increases with the disk density. 
\end{enumerate}

%
\subsection{Efficiency of Snowball Growth} \label{SECN:effi}
An efficient snowball growth phase would imply that a significant proportion of the disk's solid-mass is converted into planetesimals in this phase, so one essential condition is that $PMF_{snow}$ should be close to unity. Let's define this high-$PMF_{snow}$ condition as corresponding to $0.1<PMF_{snow}<1$. From Eq.\ref{pmfsnow}, we see that high $PMF_{snow}$ fractions are favored by high values of $H_g/H_d$. At $a=1$ AU from the star, for instance, $0.1<PMF_{snow}$ requires $H_g/H_d>40$. To a first order, $H_g/H_d$ is an indicator of the degree of turbulence in the protoplanetary disk; the larger $H_g/H_d$, the weaker the turbulence. Therefore, snowball growth is more efficient for a disk that has weaker turbulence (higher value of $H_g/H_d$).  This can be simply understood as being the condition that, for the same amount of dust mass, this mass is concentrated in a thinner, denser disk and is thus more easily accreted by planetesimals that always stay very close to the mid-plane.    

However, while the above condition is necessary, it is not sufficient. This is because, according to Eqn.\ref{pmf} (replacing $R_p$ by $R_{snow}$ and $t$ by $t_{snow}$), $PMF_{snow}$ has contributions from two sources: the snowball growth term $(R_{snow}/R_{p0})^3$, and the initial planetesimal creation term $t_{snow}/t_f$. Therefore, the efficiency of  snowball growth should be measured using both $PMF_{snow}$ and the ratio $R_{snow}/R_{p0}$. In this paper, we define the criteria for an efficient snowball growth phase as (1) $0.1<PMF_{snow}<1.0$ (or equivalently, $H_g/H_d>40$), {\it {and}} (2) $R_{snow}/R_{p0}\ge10$ (or equivalently, $t_{snow}/t_f<10^{-3}$). Note that although the efficiency of snowball growth decreases with the ratio of $t_{snow}/t_f$, the efficiency actually increases with the absolute value of $t_{snow}$, since $R_{snow}\propto t_{snow}$.

\subsection{Examples}\label{SECN:EXAMPLES}

\begin{figure}
\begin{center}
\includegraphics[width=\textwidth]{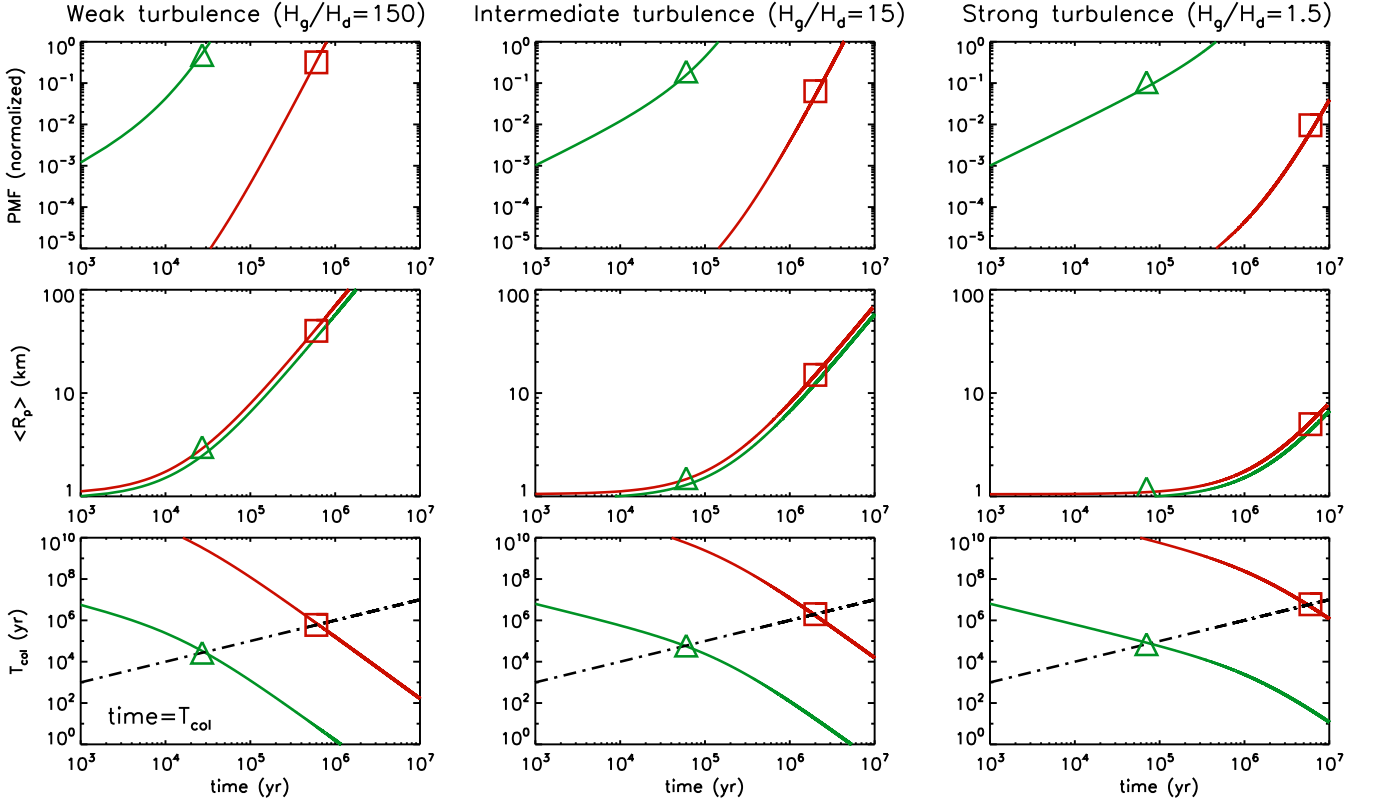}
  \caption{Planetesimal mass fraction ($PMF$), mass-weighted average radii ($\langle R_p\rangle$) and planetesimal collisional timescale ($t_{col}$) as a function of time for the cases with weak (left), intermediate (middle) and strong (right) turbulence. 
Two subcases with $t_f=10^5$ yr (in green, denoted with a triangle) and $t_f=10^{11}$ yr (in red, denoted with a square) are plotted in each panel. The dash-dot lines denote the location of $time=t_{col}$ line, so that the vertical coordinates of the triangles (or squares) indicate the location of $t_{snow}$, $R_{snow}$ and $PMF_{snow}$. N.B. the red and green lines in the 3 horizontal-middle panels which should be absolutely overlapped are deliberately plotted with a small offset. 
We see that increased $t_f$ (i.e., low planetesimal formation efficiency) allows more snowball growth to occur before collisions commence, increasing $R_{snow}$ by $\sim$ 1 order of magnitude.
In addition, we note the effect of strong turbulence in decreasing the efficiency of snowball growth, resulting in lower $PMF_{snow}$ and $R_{snow}$ occurring before planetesimal-planetesimal collisions commence.
}
\label{figpmf}
   \end{center}
\end{figure}

As shown in Eqn.\ref{pmfsnow} and discussed in the above subsection, the efficiency of the snowball growth phase depends on the ratio between the gas and dust scale heights ($H_g/H_d$) which is an indicator of the degree of disk turbulence. Here we consider three cases with weak ($H_g/H_d=150$), intermediate ($H_g/H_d=15$) and strong ($H_g/H_d=1.5$) disk turbulence, which lead respectively to $\dot{R}_p=10^{-4}, 10^{-5}$ and $10^{-6}$ $\rm km.yr^{-1}$ at 1 AU in a 1 MMSN disk. For each case, we then consider two further sub-cases with different characteristic planetesimal formation timescales (not $t_{f.ind}$): $t_f=10^5$ yr and $t_f=10^{11}$ yr and with the same initial planetesimal size of $R_{p0}=1$ km. Solving for $t_{col}$, $\langle R_p\rangle$ and $PMF$ as an explicit function of time (using Eqn.\ref{rwei}, Eqn.\ref{pmf} and Eqn.\ref{tcol}), we plot the results in Fig.\ref{figpmf}.

In Fig.\ref{figpmf}, the locations of the triangles ($t_f=10^5$ yr) and squares ($t_f=10^{11}$ yr) mark the important point at which the system transits from snowball growth to the mutual collision mode. As expected, the snowball phase is much more efficient in the weakly turbulent (high $H_g/H_d$) case. Additionally, if the initial planetesimal creation process is inefficient ($t_f=10^{11}$ yr), then the snowball phase is longer and planetesimals can grow up to $R_{snow}\sim40$ km purely via dust accretion, whereas when the creation process is efficient and $t_f=10^5$ yr, the planetesimals can only reach $\sim 3$ km before mutual collisions take over. At the end of the snowball growth phase, both cases have comparable values of $PMF_{snow}$ ($\sim35\%$ and $\sim45\%$ respectively), both of which are below our alarm value of $PMF=0.5$.

Conversely, for the highly turbulent case, snowball growth is less efficient. Although the snowball phase lasts longer than in the low turbulence case, planetesimals don't grow quickly enough to reach sizes bigger than $\sim 1.2\,$km ($t_f=10^5$ yr case) or $\sim 5\,$km ($t_f=10^{11}$ yr case). 

In summary, Fig.\ref{figpmf} demonstrates the general trend that snowball growth is more efficient when (a) planetesimal creation rates are lower (i.e., larger $t_f$), and (b) when disk turbulence is weaker (i.e., larger $H_g/H_d$). In addition, Fig.\ref{figpmf} provides a method (by finding the crossing point in the $t-t_{col}$ plane) by which $t_{snow}$, $R_{snow}$, and $PMF_{snow}$ can be found more accurately than just using the approximate expressions of Eqn.\ref{tsnow}, Eqn.\ref{rsnow} and Eqn.\ref{pmfsnow}.

\subsection{Mapping $t_{snow}$ and $R_{snow}$}\label{SECN:MAPPING}
The results of \S\,\ref{SECN:EXAMPLES} were obtained for a fixed initial planetesimal size of 1km. We now extend these results to a more general case in which the initial planetesimal size, $R_{p0}$, is allowed to vary. Such a consideration is important for two reasons: Firstly because $R_{snow}$ does not vary linearly with $R_{p0}$, and secondly because the initial planetesimal size is a poorly constrained parameter that strongly varies from one planetesimal formation scenario to the other.  

For illustrative purposes, we focus here on the most favorable case for snowball growth, i.e., weak turbulence ($H_g/H_d=150$), and vary both $R_{p0}$ and $t_f$ as free parameters. Fig.\ref{figtrsnow} shows $t_{snow}$ and $R_{snow}$ in the $R_{p0}-t_f$ plane and was generated by numerically solving the equation $t=t_{col}$. As the values of $t_{snow}$ and $R_{snow}$ plotted in Fig.\ref{figtrsnow} are accurately solved using this numerical method, they do \emph{not} rely on the approximation $\langle R_p\rangle \gg R_{p0}$ which was assumed in Eqn.\ref{tsnow} and Eqn.\ref{rsnow}. Note that in the bottom-right corner of Fig.\ref{figtrsnow}, the $R_{snow}$ contours become vertical, while the $t_{snow}$ contours become horizontal. This occurs because we suppress snowball growth whenever the PMF becomes greater than the alarm value 0.5.$t_{snow}\sim t_{tran}\sim10^6-10^7$

Not surprisingly, snowball growth is at its most efficient in the top part of the figure, where $t_f$ is long (or equivalently, the planetesimal creation rate is low).  In addition, from Eqn.\ref{rsnow} we see that $R_{snow}/R_{p0} \propto R_{p0}^{-0.25}$, so that snowball growth, measured in terms of the increase in planetesimal size, is more efficient for smaller initial planetesimals, i.e. on the left-hand side of the plot.
This result should be expected since, for a given total mass of solids, smaller initial sizes lead to larger total collisional cross-section for dust accretion.

In contrast, snowball growth gets less efficient when starting from larger initial planetesimals. However, even for $R_{p0}=100\,$km, a factor 10 increase in size (1000 in mass) is possible for low-turbulence disks with a low planetesimal formation rate ($t_f  \ge 10^{11}$ yr). Setting this factor 10 size increase as the criteria for an efficient snowball phase (see Sec.\ref{SECN:effi}), we see that the corresponding minimum value for $t_f$ increases from $\sim 10^{7}$ yr for $R_{p0}=0.1\,$km to $\sim 10^{13}$ yr for $R_{p0}=1000\,$km. 
However, another crucial constraint is that $t_{snow}$ should not exceed the time for disk dispersal (probably $\le10^7$ yr). This constraint is relevant for the following considerations: if gas is dissipated, then 
(1) both dust and planetesimals would have Keplerian velocities, and there would thus be  far less efficient dust-sweeping and no snowball growth for planetesimals, and 
(2) one would like to be able to form gas giants such as Jupiter in the Solar System, and hence require that massive solid planetary cores be able to form before disk dissipation. 
Given this additional constraint, all the solutions in the top-right part of Fig.\ref{figtrsnow} are ruled out. A factor 10 size increase in less than $10^7$ yr is thus only possible for $R_{p0}\leq100\,$km.
%

%
\section{DISCUSSION}
\subsection{Implications for Planet Formation in Single-Star Systems: The Solar System and Extrasolar Systems }\label{SECN:SS}
We emphasize that this current section of discussion (\S \ref{SECN:SS}) is based on the results of Fig.\ref{figtrsnow} for the weak turbulence case where snowball growth is most favorable.

\subsubsection{Dust to Planetesimal Transition Timescale, $t_{tran}$}
It is worth noting that in Fig.\ref{figtrsnow} we always have $PMF_{snow}\sim35\%-50\%$. As a consequence, in this case $t_{snow}$ can be interpreted as corresponding approximately to the typical timescale, $t_{tran}$, for which the solid disk goes from being dust-dominated to being planetesimal-dominated. This timescale is especially relevant for planet formation scenarios because it can be independently measured, through both protoplanetary disk observations \citep{Nat 07} and meteoritic evidence in our Solar System \citep{CW 06, Sco 07}. All these observational studies seem to agree on a transition timescale somewhere between 1 and 10Myr, i.e., $t_{snow}\sim t_{tran}\sim10^6-10^7$ yr, which is shown as the red region in Fig.\ref{figtrsnow}.

Further, we note that if $R_{snow} \sim  R_{p0}$ then $$t_{tran} \sim t_{snow} \sim {t_f},$$ whereas if $R_{snow}  \gg  R_{p0}$ then  $$t_{tran} \sim t_{snow} \ll {t_f}.$$ These relations occur because snowball growth implicitly includes two simultaneous processes, i.e. the formation of new planetesimals, plus the snowball growth of new planetesimals. As discussed in \S\,3.1, snowball \emph{growth} dominates the snowball phase only if $R_{snow}\gg R_{p0}$. This requires $R_{p0}<100 $km for the red region of Fig.\ref{figtrsnow}. 

In contrast to the above, if $PMF_{snow} \ll 0.1$ then $$t_{tran} \sim t_f  \gg t_{snow}.$$ Taking into account all these considerations, one should take care \emph{not} to confuse $t_f$ with either $t_{snow}$ or $t_{tran}$, as it may potentially be orders of magnitude different from either.

Irrespective of the precise scenario in question, we note that snowball growth provides an alternative channel through which dust can be converted into planetesimals. Moreover this can be an efficient process; even if the initial planetesimal \emph{creation} is very inefficient, one can still convert most of the solid mass into planetesimal within 1-10 Myr.

\subsubsection{Planetesimal Birth-Size}
Following the previous chapter, we take as a reference the observationally derived constraint of $t_{tran}\sim10^6-10^7$ and the additional result that, for the weak turbulence case, $t_{snow}\sim t_{tran}$. As a consequence, the most probable location for the dust-to-planetesimal transition is the red region of Fig.\ref{figtrsnow}. An important feature is that almost the whole of this red region corresponds to a $100\le R_{snow}\le1000$ km range. This implies that, for the weak turbulent case, mutual planetesimal-planetesimal collisions (or planetesimal accretion) start when planetesimals are 100-1000 km in radii, \emph{regardless of their initial size $R_{p0}$}. 

Interestingly, this implication is consistent with the result of a recent study by \citet{Mor 09}, which found that the size-distribution of asteroids in the Solar System, in particular the knee in this distribution around 100\,km, could be well reproduced if planetesimal accretion (and mutual fragmentation) starts from 100-1000 km-sized objects. This was interpreted by \citet{Mor 09} as being an indication that asteroids were born big, i.e., they emerge from their dust-to-pebbles progenitors with a size exceeding 100 km. We argue that our results could provide an alternative explanation, i.e., that 100 to 1000 km is the characteristic size for the end of snowball growth and the onset of mutual collisions, a size that might be much larger than the size $R_{p0}$ of the initial seed planetesimals. In other words, the results of Morbidelli et al. (2009) could actually provide some supporting evidence for the snowball growth, since the result $R_{snow}=100-1000$ km is robust when applying the snowball concept to parameters consistent with the Solar System. However, this inference depends on whether the weak turbulence case is consistent with the Solar System. 

As mentioned in $\S$1, a form of snowball growth was taken into account in the simulations of \citet{Mor 09} (see their Fig.6b). However, there are important differences between their work and this present study. First of all, dust sweeping was only considered for large ($\geq 100\,$km) seed planetesimals, implicitly neglecting its possible effect on smaller 
initial bodies. Another, perhaps more important difference lies in the respective models adopted for dust sweeping. \citet{Mor 09} argue that it proceeds in a fast ``runaway'' mode -- thus making it end much earlier -- because the relative velocity between the planetesimals and the dust ($v_{vel}\sim75 \rm ms^{-1}$) is smaller than the planetesimals' escape velocity ($v_{esc}\sim100 \rm ms^{-1}$ for a 100 km size), and hence the dust is gravitationally focused onto the larger planetesimals. In this study we ignore runaway growth, even for large planetesimals, because we implicitly assume that the coupling of the dust to the gas is strong enough to prevent solid grains from being gravitationally deflected onto the planetesimals according to the usual gas-free approximation. The issue of whether or not the runaway mode is significant for snowball growth would require an investigation using high resolution coupled N-body and hydrodynamic models that exceed the scope of this paper.

In summary, we believe that the interpretation of the present-day asteroid size distribution as a proof of their large initial sizes \citep{Mor 09} needs further investigation in the light of the results presented here on the importance of dust sweeping for early planetesimal growth. At the very least, snowball growth opens up the possibility that various conditions with different combinations of $R_{p0}$ and $t_f$ could all potentially reproduce the asteroids' size-distribution, to the point that the solution may be highly degenerate. In fact, recent work by \citet{Wei 10} has already shown that planetesimals with birth-size ($R_{p0}$) of 0.1 km are also a viable initial condition for a model which can well reproduce the asteroids' size-distribution.   

Further work would be required, using a detailed numerical collision-and-fragmentational model (E.g. as implemented in \citet{Mor 09} or \citet{Wei 10}), to understand whether the observed size distribution in the asteroid belt can at all constrain the range of $R_{p0}$ and $t_f$ in the early Solar System. I.e. it may be that certain ranges of $R_{p0}$ and $t_f$ corresponding to sub-regions of the red regime of Fig.\ref{figtrsnow} may give rise to size distributions that are better able to reproduce the observed size distribution in the asteroid belt, while other regions can be excluded.

However, even so, we would still be far from arriving at a final answer, as several important questions remain unaddressed, including 
(1) What is the distribution of sizes with which planetesimals are initially created?
(2) How are the planetesimal births distributed in time and space (linear or nonlinear)?
(3) To what degree should runway growth and gravitational focusing be considered in the snowball growth model?  
All of these factors can change the dimension of the red regime of Fig.\ref{figtrsnow}, and constraining any of these factors will require further detailed studies of planetesimal formation itself. Last but not least, in addition to studies of the formation of asteroids and planetesimals, we may also obtain constraints on planetesimal birth sizes through other observations, such as those of debris disk around A-type and G-type stars \citep{KB 10}.

\subsubsection{Planetesimal Birth-Rate}
\begin{figure}
\begin{center}
\includegraphics[width=\textwidth]{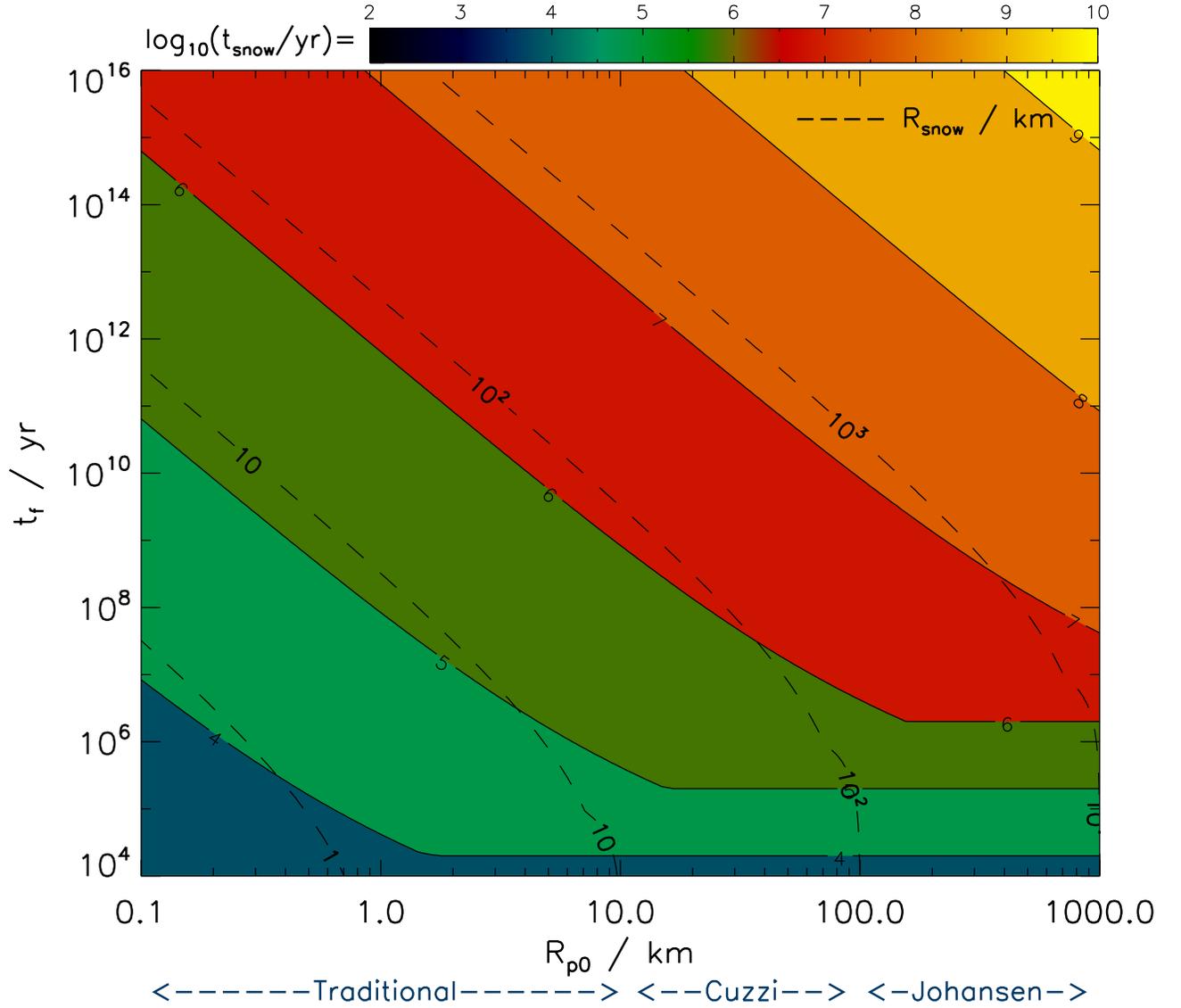}
  \caption{$t_{snow}$ (solid contours filled with colors) and $R_{snow}$ (dashed contours) in the $R_{p0}-t_f$ plane at 1 AU for 1 MMSN with $H_g/H_d=150$ (weak turbulence) in a single star system. The $R_{p0}$-coordinate is schematically divided into three size ranges for the initial planetesimals, forming via traditional models (Goldreich \& Ward 1973; Weidenschilling \& Cuzzi 1993) and via two new models which involve turbulence (Johansen et al. 2007; Cuzzi et al. 2008). 
Snowball growth is most efficient in the top-left corner of the plot, where planetesimals of initial size 0.1 km can grow via direct dust accretion (snowball growth) to $\sim$ 100 km within $10^6$ yr, before planetesimal-planetesimal collisions commence. 
Note that, for $t_{snow}$ in the $\sim10^6-10^7$ yr range (the typical life-time of a protoplanetary disk), we obtain $100<R_{snow}<1000$ km \emph{regardless} of the initial value $R_{p0}$.
}
\label{figtrsnow}
   \end{center}
\end{figure}

The results shown in Fig.\ref{figtrsnow} also allow us to place constraints on the birth rate of planetesimals or their formation efficiency for each possible formation scenario. In addition, they might help overcome some difficulties that some of these different formation scenarios encounter. 

Recently two new models \citep{Joh 07, Cuz 08} have shown that large planetesimals can form directly from the concentration of small solid particles in the turbulent disk.
The numerical simulation by \citet{Joh 07} show that if the feedback of the solid particles within the gas is considered then the concentration of solid particles (which is most efficient for particles of $\sim 50$ cm radial size) can grow and be maintained long enough (the so-called streaming instability, \citet{YG 05}), for the formation of very large planetesimals (typically 100-1000 km). Within this framework, i.e., with such large initial planetesimals, our results (Fig.\ref{figtrsnow}) show that the phase of snowball growth would be inefficient. In such a scenario, the only way to convert a large amount of dust to planetesimals within 1-10Myr is through a very effective planetesimal formation process. 
Nevertheless, one must be careful and note that Fig.\ref{figtrsnow} ignores the runaway growth mode which might be relevant to such large planetesimals. If a runaway mode is included, then the snowball timescale, $t_{snow}$,  given in Fig.\ref{figtrsnow} should be considered as a upper limit. However, as discussed in \S\,4.1.2, the degree to which a runaway mode should contribute must rely on future studies with high resolution coupled N-body and hydrodynamic models. Moreover, one should also note that disk turbulence is much more intense in \citet{Joh 07} than that in Fig.\ref{figtrsnow} of this paper. As in \citet{Joh 07}, the disk viscosity coefficient is adopted $\alpha\sim10^{-3}$, leading to $H_g/H_d$ close to unity according to Eqn.108 of \citet{Arm 10} and thus inefficient snowball growth would result. It would be interesting to investigate the size and efficiency with which planetesimals can be formed via the mechanism of \citet{Joh 07} but with weaker turbulence.

The other large-planetesimal model \citep{Cuz 08} also relies on turbulence which can generate vortices to trap small particles \citep{HK 10} of chondrule-size (typically mm-cm). \citet{Cuz 08} showed that mm-sized particle can be sporadically concentrated by disk turbulence, forming large and gravitationally bond clumps, which can potentially shrink to solid planetesimals roughly 10-100 km in radius. However, based on the model of \citet{Cuz 08}, \citet{Cha 10} and \citet{Cuz 10} investigated the planetesimal formation rate in details and found that it should be very low in the Solar System: at 1 AU for a typical MMSN disk, the time it would take to convert most dust into planetesimals, $t_f$, would be over $\sim10^{10}$ yr. Snowball growth might help in solving this problem by creating a second channel by which dust can be converted into planetesimals. Fig.\ref{figtrsnow} shows that, for the case of weak turbulence (this turbulence level in Fig.\ref{figtrsnow} is compatible with the weak turbulence adopted in \citet{Cuz 08}) , the time to convert dust into planetesimals is close to $t_{snow}$, which is much shorter than $t_f$. In fact, $t_{snow}$ can be within the disk life-time (1-10 Myr) even for $t_f\ge10^{10}$ yr. In other words, snowball growth would fully dominate in terms of transforming dust into planetesimals.

Beside the two new models described above, there are the two traditional models of planetesimal formation described in \S\,1. One is the mutual sticking scenario \citep{WC 93, Wei 97}, in which planetesimals form by pair-wise collisions of small particles. The other is the instability scenario \citep{Saf 69, GW 73, YS 02}, which suggests that planetesimals form via gravitational instabilities in a dense particle sub-disk near the midplane of the protoplanetary disk. Both scenarios remain debated. 

As mentioned in our introduction, the collisional scenario is challenged by the well-known ``meter-barrier", while for the instability scenario it has not yet been established whether a disk of solid particles can become dense enough and dynamically cold enough to trigger instability.  While these issues go well beyond the scope of this work, the present results show that, for both scenarios, snowball growth might represent an interesting way to allow planetesimal growth to happen despite these difficulties. Indeed, in both models planetesimals are thought to form with a radius of 0.1-10 km. For this size range, Fig.\ref{figtrsnow} shows that snowball growth could convert most of the dust mass into planetesimals even if formation rates are very low (large $t_f$). This means that only a few "lucky seeds" are needed in order for planetesimal growth to proceed. 
Within the framework of these two scenarios, these seeds could either be a few lucky pre-planetesimals that have overcome the ``meter barrier" for some reasons \citep{Bra 08, Joh 08, TW 09, Wet 09} or kilometer-sized objects that were formed in some isolated location where the critical condition for gravitational instability just happens to be satisfied \citep{GW 73}. Therefore, snowball growth lowers the requirements for those two traditional scenarios: direct planetesimal formation from dust could be very inefficient and still allow planetesimal growth to occur. This hypothesis remains to be quantified: future work should focus on deriving the planetesimal formation rates within the frame of those two traditional scenarios, to see whether they can meet these ``lowered'' requirements.

\subsection{Implication for Planet Formation in Close Binaries}
The context in which snowball growth might find its most interesting application is that of planet formation in binaries. Recent studies have shown that one of the major problems for planet formation in close binary systems, such as $\alpha$ Centauri, is the intermediate stage of planet formation, i.e, the mutual accretion of  km-sized planetesimals to form larger planetary embryos or cores \citep{The 06, Hag 09a, Hag 09b}. The companion's gravitational perturbation, coupled to gas drag, may excite large relative velocities between the planetesimals, leading to disruptive collisions which inhibit their mutual accretion \citep{The 08, The 09, Par 08, Mar 09, Xie 10}. Recently, some mechanisms have been found to be somewhat helpful in solving this problem \citep{XZ 08, XZ 09, PL 10},  but several problems remain to be overcome.
  
 A possible solution to these problems would be to let planetesimal collisions only begin when large ($\geq 50-100\,$km) objects are present in the system. \citet{The 08} have indeed shown that the gravity of such large planetesimals is strong enough for them to survive the high-speed collisions induced by differential gas drag. One obvious way to bypass this problematic kilometer size range would be to have planetesimals that are "born big", be it by the \citet{Joh 07} or the \citet{Cuz 08} scenario. However, the issue of how these planetesimal formation scenarios might work in the very specific context of binaries has not yet been addressed. 

Snowball growth offers an attractive alternative way to bypass the kilometer size range, \emph{regardless} of the initial size at which planetesimals appear in the disk, because it predicts that mutual planetesimal impacts will become important only by the time large objects have formed. However, it is not possible to directly apply the results of the previous sections, derived for single stars, to the close binary case. In this section, we address the issue of how snowball growth proceeds in the specific context of double stars.

\subsubsection{An Example: \emph{$\dot{R}_p$ and $t_{col}^b$} in $\alpha$-Centauri}
For the sake of clarity we consider a disk with a 1$\times$MMSN surface density and set $a=1$ AU, and then investigate how snowball growth may occur in binary star systems, taking as a representative example the case of $\alpha$ Centauri AB, for which the binary separation is $a_B=23.4$ AU and the eccentricity $e_B\sim$ 0.52 \citep{Pou 02}.

The main difference between the single star case and the binary case is that in the binary case there is a much higher relative velocity between the dust and the planetesimals ($v_{rel}$). For such a highly eccentric case, we can safely neglect the small component of $v_{rel}$ due to the differential Keplerian velocities between the gas and the planetesimals (because the gas is pressure supported)  and basically consider two cases, depending on the unknown angle $i_B$ between the circumprimary disk plane and the binary's orbital plane:

\begin{enumerate}
\item \emph{The coplanar binary case}, where the binary orbit and the gas disk are in the same plane. In this case, to a first approximation, the average eccentricity of planetesimals ($e_p$) is equal to their forced eccentricity ($5e_Ba/4a_B$) {\footnote{In reality, the equilibrium eccentricity can depart from this purely dynamical forced value, as it also depends on gas drag (see Eqn.24 of \citet{Par 08}).  However, we ignore this refinement for the order-of-magnitude feel of the present discussion.}}, and thus $v_{rel}\sim e_pv_k\sim(5a/4a_B)e_Bv_k$, leading to $v_{rel}\sim840 \ \rm ms^{-1}$ for the $\alpha$ Cen case. 
This $v_{rel}$ value, and thus also the snowball growth rate, is roughly one order of magnitude higher than in the single star case. We then obtain $\dot{R}_p\sim10^{-3}\rm km.yr^{-1}$ in a weakly turbulent disk with $H_g/ H_d=150$. 

Note that we here implicitly assume that the gas disk is axisymmetric (circular) in the circumprimary midplane. This is a rather crude approximation as it ignores the reaction of the gas disk to the binary perturber. Nevertheless, as a first order of approximation, we believe it is both reasonable and convenient for our semi-analytical study in this paper. In fact, if the reaction of the gas disk is considered, $v_{rel}$ would probably be even larger, as shown by Paardekooper et al. (2008). Therefore, the $v_{rel}$, as well as the snowball growth rate $\dot{R}_p$ derived in our simplified axisymmetric gas disk model should be treated as a \emph{lower} limit.

\item \emph{The inclined binary case}, where the binary orbital plane is tilted by an angle ($i_B$) relative to the gas disk plane. In such a case, as simulated by \citet{Lar 96}, if the Mach number of the gas disk is not too high, the gas disk can maintain its structure and undergo a near rigid precession. As a planetesimal also undergoes a similar precession (regarding the binary orbital plane as the reference plane) but with a different rate, the relative angle between the gas disk plane and planetesimal orbital plane will remain approximately within the range $0-2i_B$. Taking the medium value $i_B$ as the average, we then have $v_{rel}\sim i_Bv_k\sim520 (i_B/1^\circ)\ \rm ms^{-1}$. 
Taking into account this $v_{rel}$, as well as the fraction of time that a planetesimal spends moving within the dust-disk, $2H_d/\pi ai_B$, then the snowball growth rate for the inclined case is $\dot{R}_p\sim10^{-5}\rm km.yr^{-1}$, which is independent of both $i_B$ and the turbulence factor $H_g/H_d$.

Note that, for the inclined case, we have implicitly assumed that the vertical excursion of the planetesimals is higher than the dust disk thickness, i.e., $i_B > 0.05 (H_d/H_g) (a/AU)^{0.25}$. However, even for a fully turbulent disk ($H_d/H_g=1$), this condition is easily met as long as $i_B>3^\circ$. In addition, we also implicitly assume for the inclined case that $v_{rel}$ is dominated by $i_B$ rather than $e_B$, which requires $520 (i_B/1^\circ)>840$ (i.e. comparing $v_{rel}$ in the the inclined and coplanar cases), i.e., $i_B>1.6^\circ$ {\footnote{This small value is due to the fact that we are considering a region (1\,AU) close to the primary, where the forced eccentricity, which decreases as $1/a$, is low whereas the forced inclination, which is independent of the semi-major axis, stays at a high value}}. Given the above considerations, we shall consider that the inclined case discussed here corresponds to cases with $i_B>3^\circ$. Otherwise it should be treated as the coplanar case. 

\end{enumerate}

The collision timescale, $t_{col}$, among planetesimals is also different in the binary case. The relation given in Eqn.\ref{tcol}, which assumes an unperturbed disk with equipartition between in and out-of-plane velocities, breaks down in binary systems. Here we use the empirical scaling law given by \citet{Xie 10} (see Eqn.6 and Eqn.7 in their paper for details) to estimate the planetesimal collisional timescale in binary star systems as 
\begin{equation}
t_{col}^b\sim(0.02+2i_B) t_{col}^s,
\label{tcolb}
\end{equation}
 where $t_{col}^s$ is the collisional timescale for single star systems as given by Eqn.\ref{tcol}. 
 \\

\subsubsection{An Example: The Snowball Phase in $\alpha$-Centauri}

\begin{figure}
\begin{center}
\includegraphics[width=\textwidth]{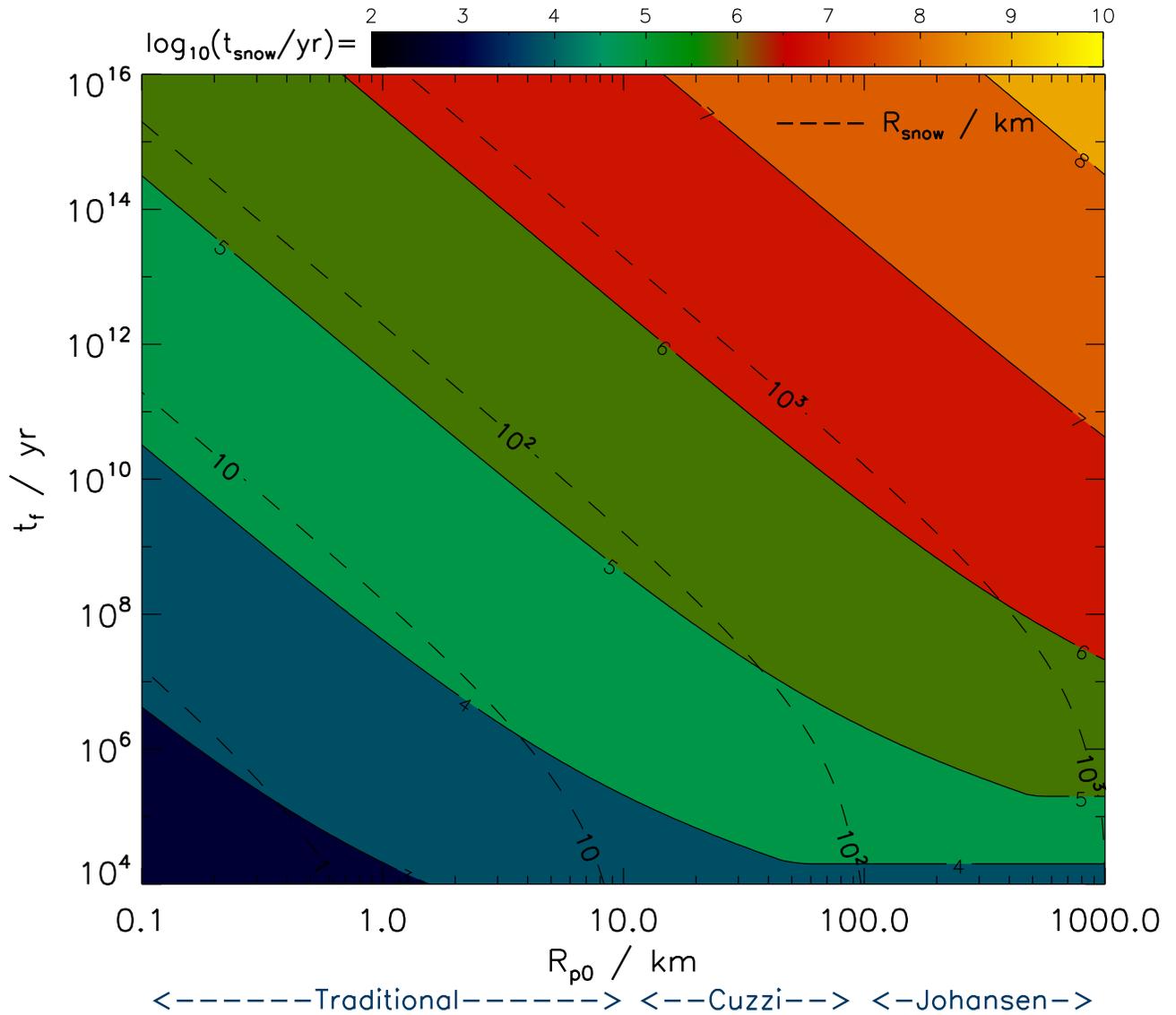}
  \caption{Same as Fig.\ref{figtrsnow} but for the coplanar binary case ($i_B=0$) and $H_g/H_d=150$. The binary system parameters are those of $\alpha$ Centauri. $R_{snow}$ is close to its value for the single star case Fig.\ref{figtrsnow}, but $t_{snow}$ is about one order of magnitude smaller, implying that snowball growth is efficient even if the disk life-time in close binary systems is an order of magnitude shorter than in single-star systems.}
  \label{figtrsnow2}
   \end{center}
\end{figure}

\begin{figure}
\begin{center}
\includegraphics[width=\textwidth]{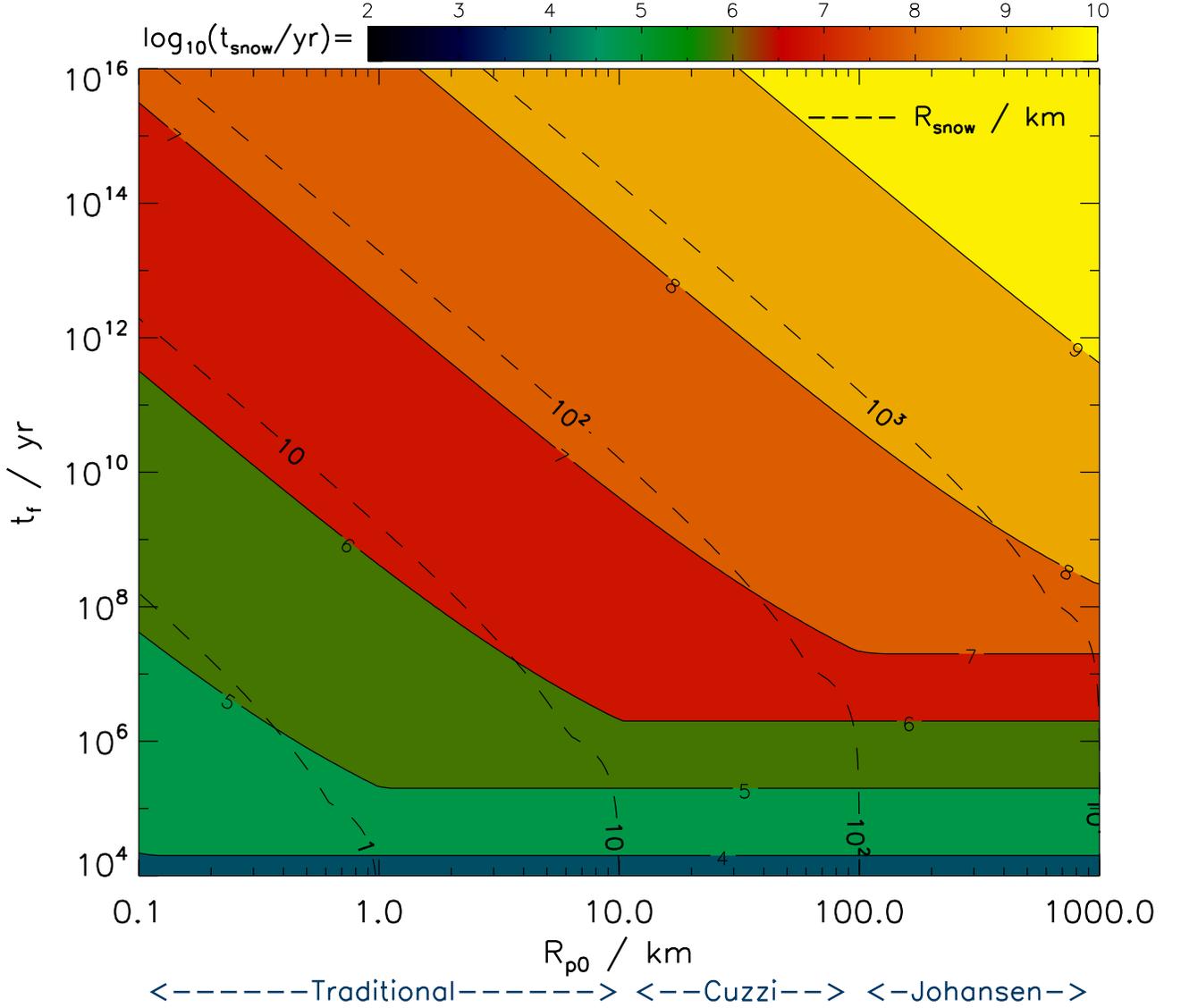}
  \caption{Same as Fig.\ref{figtrsnow2} but for the inclined binary case with $i_B=10^ \circ$.  The binary system parameters are those of $\alpha$ Centauri. Note that the present results are independent of the turbulence factor $H_g/H_d$.  Compared to the single star case of Fig.\ref{figtrsnow}, $R_{snow}$ is smaller by $\sim30\%$, but $t_{snow}$ is larger by about 1 order of magnitude. If, for example, one expects planetesimals to be born small ($R_{p0}<10$ km) but want them to have a large $R_{snow}$ (10-100 km) in order that they can survive high-speed collisions, then the disk life-time should be no less than $\sim10^6$ yr.}
   \label{figtrsnow3}
   \end{center}
\end{figure}

With the modified $\dot{R}_p$ and $t_{col}^b$, we can derive $t_{snow}$, $R_{snow}$ and $PMF_{snow}$ as in the single star case. Results are plotted in 
Fig.\ref{figtrsnow2} and Fig.\ref{figtrsnow3} for the coplanar and inclined binary cases respectively.

\begin{enumerate}
\item \emph{Coplanar case}. Compared to the single star case of Fig.\ref{figtrsnow}, we obtain slightly smaller, but still relatively large values of $PMF_{snow}$, i.e., $\sim7\%-50\%$. $R_{snow}$ is little changed, but $t_{snow}$ is about one order of magnitude smaller (Fig.\ref{figtrsnow2}). These high $PMF_{snow}$ and low $t_{snow}$ values imply that snowball growth can be efficient, even if the disk life-time in such close binaries could be an order of magnitude shorter than in single-star systems \citep{XZ 08, Cie 09}. Note that, as for the single star case, results depend on the disk turbulence, which is controlled, to a first approximation, by $H_g/H_d$. Here, we have assumed a value of $H_g/H_d=150$, corresponding to weak turbulence.

\item \emph{Inclined binary case}.  $PMF_{snow}$ are here higher than in the coplanar case, being close to $50\%$ throughout the $R_{p0}-t_f$ plane of Fig.\ref{figtrsnow3} (In fact, $PMF_{snow}>50\%$, but we stop the evolution of the system when $PMF_{snow}$ reaches the alarm value 0.5 as mentioned at the end of section 2.2). Compared to the single star case of Fig.\ref{figtrsnow}, $R_{snow}$ is smaller by $\sim30\%$, but $t_{snow}$ is larger by about 1 order of magnitude (more precisely, by a factor of $\sim$7). This implies that the disk lifetime has to be relatively long in order for snowball growth to be efficient. As an example, if planetesimals are born small ($R_{p0}<10$ km), then in order for them to reach $R_{snow}\sim 50-100\,$km, i.e., values large enough to survive the high-speed collisions induced by the companion, then the disk life-time should be no less than $\sim10^6$ yr. Note that, as $\dot{R}_p$ is independent of $H_g/H_d$ (see \$\,4.2.1), the result is here independent of the level of disk turbulence.
\end{enumerate}

The results for both cases considered show that snowball growth might indeed be an attractive alternative to solving the planet-formation-in-binaries dilemma, i.e., the high impact velocities that prevent the mutual accretion of kilometre-sized bodies. It's main appeal is that it provides a much safer way for planetesimals to reach sizes in the 50-100\,km range, after which they are large enough to be protected from destructive collisions. The only problem lies with the growth timescales, in particular for the inclined binary case for which it might take as long as $10^{6}\,$yr to reach the safe 100\,km size, by which time the gas disk might have vanished. However, this issue might not be as dramatic as it might appear because, even if 50-100\,km bodies have not been formed by the time the gas disk is dispersed, the environment could be favorable to accretion anyway. Indeed, \citet{XZ 08} have shown that, during this gas dispersion phase, planetesimal orbits get re-phased so that, by the time the gas has dispersed, relative velocities amongst them are very low and accretion friendly. This means that, even if snowball growth is switched off (because there is no dust sweeping when grains stop to follow gas streamlines), planetesimal growth can now proceed through the classical mutual-accretion channel.

We note that the 2-D investigation performed by \citet{PL 10} demonstrated that planetesimals can grow to $\sim$50-100 km size with the help of dust accretion. Our own investigation extends this to the 3-dimensional case, demonstrating that even though 3-D collision rates are significantly lower, dust accretion (or snowball growth) can still help planetesimals to overcome the ``km-barrier" for planetesimal accretion (or growth) in close binary systems, such as $\alpha$ Centauri.

\subsubsection{limitations}
One possibly problematic assumption of our snowball growth model regards the efficiency of dust accretion onto planetesimals, parameterized by $E_d$ defined in Eqn.\ref{sweep}, for which we have implicitly assumed that it is not affected by the high $v_{rel}$ values reached in the binary environment. Although recent laboratory experiments \citep{TW 09} showed that small (mm-sized) particles can be efficiently accreted by large objects in a high-speed collision, their test speeds were only of a few 10 ms$^{-1}$, which is only relevant for the snowball growth in a single star system with $v_{rel}\sim 75$ ms$^{-1}$. In close binary systems, such as the $\alpha$ Centauri shown above, $v_{rel}$ could be as high as 500-5000 ms$^{-1}$, which is beyond the limit of the test speed in any existing laboratory experiment. Therefore, the issue of whether dust can be accreted onto small planetesimal seeds ($R_{p0}=0.1-1$ km) for such large relative velocity remains unresolved. However, at the very least, it seems reasonable to assume that the efficiency of dust accretion onto planetesimals will be \emph{less} affected by high velocities than would the efficiency of mutual planetesimal accretion.

\section{CONCLUSION AND PERSPECTIVES}
We have investigated the transitional phase of planet formation,  dubbed ``snowball growth phase'', in which planetesimals are progressively emerging from the dust disk but are not yet numerous enough to enter the phase of mutual planetesimal-planetesimal collision. In this snowball phase, the planetesimals move on Keplerian orbits and grow purely through the direct accretion of sub-cm sized dust which is entrained with the gas in the protoplanetary disk and moves with sub-Keplerian velocity.

Using a simplified model in which the planetesimals are progressively created from the dust disk, we find that:

\begin{enumerate}
\item In single-star systems, snowball growth can be a significant mechanism if the following conditions are met:
\begin{itemize}
 \item The dust disk must be thin and dense, i.e., the gas disk must be weakly turbulent (see \S\,\ref{SECN:EXAMPLES});
 \item The planetesimal creation timescale, defined by the time to convert half the total dust mass into planetismals by the planetesimal-formation process alone, must be long (roughly $t_f>10^7$ yr - see \S\,\ref{SECN:EXAMPLES});
 \item Planetesimals must be born (relatively) small: if planetesimals are born larger than 100km, then (ignoring the runaway growth mode) snowball growth will be inefficient or completely absent (see \S\,\ref{SECN:MAPPING}). 
\end{itemize}
\emph{If} these conditions are met, then we find that

\begin{itemize}
 \item  Size growth can be significant: 2 orders of magnitude in radius (6 orders in mass), before mutual planetesimal collisions take over.  In addition, snowball growth can be fast enough for it to happen within the typical (1-10Myr) lifetime of a protoplanetary gas disk (see \S\,\ref{SECN:MAPPING}).  
 \item  Snowball growth can be the dominant mode for the transformation of dust into planetesimals. Except for highly efficient and fast planet-formation scenarios, dust is preferentially accreted onto already formed kilometre-sized bodies rather than consumed for forming new ``initial" planetesimals (see \S\, 4.1.1).
\item Applying the snowball growth model to a fiducial MMSN disk for the Solar System and assuming weak turbulence, then the turnover from snowball to mutual collisions is likely to occur when planetesimals reach a size of 100-1000 km, \emph{irrespective} of their birth size (E.g. 0.1 - 100km). This result, i.e., that mutual impacts become significant only when large bodies have formed, may provide an alternative explanation to the size distribution of asteroids (\citet{Mor 09}, see \S\,4.1.2).
 \item Snowball growth could help overcome some still unsolved problems inherent to some planet formation scenarios. More precisely, it lowers the requirements on the planetesimal formation mechanism itself, by allowing planetesimals to grow, and dust to be transferred onto them, even if the mechanism by which they are formed is very inefficient (see \S\,4.1.3).  
\end{itemize}

\item  For close binaries, snowball growth offers a way to overcome one major problem for planet formation in binaries, i.e., the high impact velocities that prevent kilometer-sized planetesimals from accreting each other. Indeed, this dynamically excited environment, which is hostile to mutual planetesimal accretion, is on the contrary favorable to growth by dust sweeping. If efficient enough, snowball growth allows planetesimals to grow large enough, 50-100\,km, to be protected from mutually destructive impacts. Two main cases can be distinguished:
\begin {itemize}
\item In the coplanar binary case ($i_B<3^\circ$), snowball growth is at its most efficient (a little more efficient than the single star case, if one assumes the same level of disk turbulence) but remains sensitive to disk turbulence, i.e. weak turbulence is still required for efficient snowball growth (see Fig.\ref{figtrsnow2} and \S\,4.2.2).   
\item In the case of an inclined binary ($i_B>3^\circ$), snowball growth can still be efficient, although slightly less efficient than in the case of a single star with weak disk turbulence($H_g/H_d=150$), with the added advantage that it is independent of disk turbulence (see Fig.\ref{figtrsnow3} and \S\,4.2.2).  

\end{itemize}  
\end{enumerate}

Let us stress again that our model is highly simplified and that the present work, whose main objective was to identify a new and potentially significant mechanism, should be considered as a first step towards more detailed studies. In future work, the modeling of snowball growth will need to be improved in a number of aspects. The most important improvement will be to have a self consistent model of the planetesimal size evolution under the coupled effect of snowball growth and mutual accretion, instead of the simple analytical growth law assumed here. This requires the development of a global statistical particle-in-a-box model, in the spirit of those of \citet{Wei 97} or \citet{Mor 09}, spanning a wide range in sizes and incorporating all possible outcomes for dust-planetesimal and planetesimal-planetesimal collisions.  Other improvements which could be made to the model include, (i) the role of gravitational focusing of dust onto planetesimals, in particular to what extent it is (or is not) hampered by the action of gas, and (ii) the changes to $E_d$ arising from the physics of high velocity impacts of dust onto large bodies, especially in the specific case of close binaries where $v_{rel}$ may reach values of a few 100m.s$^{-1}$.


\section{Acknowledgments}
 This work was supported by the National Natural Science Foundation of China
  (Nos.10833001, 10778603 and 10925313), the National Basic Research Program of China(No.2007CB814800), NSF with grant AST-0705139 and 0707203, NASA with grant NNX07AP14G and NNX08AR04G, W.M. Keck Foundation, Nanjing University and also University of Florida. J.-W. Xie was also supported by the China Scholarship Council. M. J. Payne was also supported by the NASA Origins of Solar Systems grant NNX09AB35G and the University of Florida's College of Liberal Arts and Sciences.
%

%

%

\clearpage


\clearpage

\clearpage

\clearpage

\begin{table}
 \begin{center}
  \caption{List of Notions.\label{tbl-1}}
  \begin{tabular}{@{}llllllllllrrr@{}}
  \hline
\tableline \\
 &       Variables                   &       \ \ \ \ \ \ \ \ \ \ \ \ \ \ \ \ \ \ \ \ \ \ \ \ \ \ \ \           Meaning          & &                    Definition \\ \\
\tableline
& $a_{B}$.....................        &      Semimajor axis of the orbit of binary system                   & &              First paragraph of \S\,4.2.1 \\[2pt]
& $a_{ice}$...................      &     Ice boundary (a semimajor axis of ice condensation)          & &              Just behind Eqn.(\ref{grate}) \\[2pt]
& $E_d$.....................             &     Dust accretion efficiency factor                                          & &                 Eqn.(\ref{sweep}) \\[2pt]
& $e_{B}$.....................       &      Eccentricity of the orbit of binary system                         & &              First paragraph of \S\,4.2.1 \\[2pt]
& $e_p$......................            &      Planetesimal average orbital inclinatio                             & &               Second paragraph of \S\,4.2.1 \\[2pt]

& $f_d$......................             &     Enhancement factor of $\Sigma_d$ from a nominal disk (MMSN)     & &      Just behind Eqn.(\ref{grate}) \\[2pt]
& $f_{ice}$....................       &     Enhancement factor of $\Sigma_d$ due to ice condensation     & &        Just behind Eqn.(\ref{grate}) \\[2pt]

& $H_d$.....................            &     Scale height of the dust-disk                                                & &                 Eqn.(\ref{sweep}) \\[2pt]
& $H_h$....................            &     Scale height of the gas-disk                                                  & &                 Eqn.(\ref{grate}) \\[2pt]
& $i_{B}$......................         &      Inclination of the orbit of binary system                            & &              Second paragraph of \S\,4.2.1 \\[2pt]
& $i_p$......................            &      Planetesimal average orbital inclinatio                             & &                Eqn.(\ref{tcol}) \\[2pt]
&  $M_{p0}$..................     &     Initial individual mass of the planetesimals                       & &                  Eqn.(\ref{rate}) \\[2pt]
& $\dot{M}_p$....................  &     Planetesimal mass growth rate via dust accretion              & &                 Eqn.(\ref{sweep}) \\[2pt]
& ${M}_p$....................       &     Planetesimal mass                                                                & &                 Eqn.(\ref{sweep}) \\[2pt]
&            $N$......................  &       Surface density of the planetesimals                                & &                  Eqn.(\ref{rate}) \\[2pt]
& $n$.......................                &      Planetesimal volume density                                                & &                Eqn.(\ref{tcol}) \\[2pt]
& $PMF$................         &      Planetesimal mass fraction in the total solid mass               & &                Eqn.(\ref{pmf}) \\[2pt]
& $PMF_{snow}$.........  &     $PMF$ at the end of snowball phase                                & &                Eqn.(\ref{rsnow}) \\[2pt]
& $R_{p0}$...................      &      Initial individual radial size of the planetesimals                 & &                Third paragraph of \S\,2.1 \\[2pt]
& $R_p$.....................            &     Planetesimal radii                                                                & &                 Eqn.(\ref{sweep}) \\[2pt]
& $\dot{R}_p$.....................   &     Planetesimal radial growth rate via dust accretion               & &                 Eqn.(\ref{grate}) \\[2pt]
& $\left<R_p\right>$.................. &     Mass-weighted average radii of the already-formed planetesimals         & &       Eqn.(\ref{rwei}) \\[2pt]
& $R_{snow}$............... &     $\left<R_p\right>$ at the end of snowball phase              & &                Eqn.(\ref{rsnow}) \\[2pt]

& $t_{col}$....................      &      Timescale for the onset of planetesimal-planetesimal collision               & &                Eqn.(\ref{tcol}) \\[2pt]
&  $t_{col}^b$....................  &      $t_{col}$ in binary star systems                                              & &                Eqn.(\ref{tcolb}) \\[2pt]
&  $t_{col}^s$....................   &      $t_{col}$ in single star systems                                              & &                Eqn.(\ref{tcolb}) \\[2pt]
& $t_{f.ind}$................       &     Formation timescale of individual planetesimals               & &                 Second paragraph of \S\,2.1 \\[2pt]
& $t_{f}$......................        &     Characteristic planetesimal formation timescale                & &                 Eqn.(\ref{tform}) \\[2pt]
& $t_{snow}$................  &      Duration of snowball phase                                              & &                Eqn.(\ref{tsnow}) \\[2pt]
& $t_{tran}$.................      &      Transition timescale from dust-dominated to planetesimal dominated        & &    First paragraph of \S\,4.1.1 \\[2pt]
& $v_{esc}$...................     &      Planetesimal escape velocity                                            & &              Second paragraph of \S\,4.1.2 \\[2pt]
& $v_{k}$......................        &      Local Keplerian velocity                                                      & &                 Eqn.(\ref{vrel}) \\[2pt]
& $v_{rel}$...................       &     Relative velocity between planetesimal and dust                & &                 Eqn.(\ref{sweep}) \\[2pt]
%
%
& $\Delta V$...................    &      Planetesimal average relative velocity                             & &                Eqn.(\ref{tcol}) \\[2pt]
&  $\epsilon_p$...................... &    Planetesimal formation efficiency factor                            & &                  Eqn.(\ref{rate}) \\[2pt]
& $\rho_*$......................       &     Planetesimal internal density                                                & &                 Eqn.(\ref{grate}) \\[2pt]
&  $\Sigma_d$.....................  &                   Dust surface density                                             & &                  Eqn.(\ref{rate}) \\[2pt]

  \hline
\tableline 
\end{tabular}
\tablecomments{The table list is on the alphabetical order of the initial of the variable name. Several variables in Greek are listed at the end of the table.}
\end{center}
\end{table}

\appendix
\section{mass-weighted average radii: $\langle R_p\rangle$}
As we assume a linear planetesimal formation rate (see Eqn.\ref{rate}) and a linear growth rate (see Eqn.\ref{grate2}), thus, at a given time $t$, the distribution of the planetesimals' radii will be given by $R_{p0}, R_{p0}+dr, R_{p0}+2\times dr, ... , R_{p0}+i\times dr, ... , R_{p0}+(n-1)\times dr$, where $dr$ is the radii increase during a time interval of $t_{f.ind}$, i.e., $dr=\dot{R}_pt_{f.ind}$, and $n$ denotes the total number of planetesimals. 
The total mass of these planetesimals is 
\begin{eqnarray}
M_{tot} &=& \frac{4\pi}{3}\rho_*\sum(R_{p0}+i\times dr)^3 \nonumber \\
&=&\frac{4\pi}{3}\rho_*(R_{p0}^3n+3R_{p0}^2dr\sum i +3R_{p0}dr^2\sum i^2  +dr^3\sum i^3).
\label{Mtot}
\end{eqnarray}
We define the mass-weighted average radii, $\langle R_p\rangle$, which satisfies 
\begin{eqnarray}
M_{tot}=\frac{4\pi}{3}\rho_* \langle R_p\rangle^3n.
\label{rdefine}
\end{eqnarray}
Comparing Eqn.\ref{Mtot} and Eqn.\ref{rdefine}, we have  
\begin{eqnarray}
\langle R_p\rangle &= &(R_{p0}^3+\frac{3}{n}R_{p0}^2dr\sum i+\frac{3}{n}R_{p0}dr^2\sum i^2+\frac{1}{n}dr^3\sum i^3)^{1/3}\nonumber \\
&=&[R_{p0}^3+\frac{3}{2}R_{p0}^2(n-1)dr+R_{p0}(n-1)(n-\frac{1}{2})dr^2+\frac{1}{4}n(n-1)^2dr^3]^{1/3}.
\label{r1}
\end{eqnarray}
Considering $$(n-1)dr\sim(n-\frac{1}{2})dr\sim ndr\sim \dot{R}_pt,$$ then Eqn.\ref{r1} can be rewritten as 
\begin{eqnarray}
\langle R_p\rangle&\sim&[R_{p0}^3+\frac{3}{2}R_{p0}^2(\dot{R}_pt)+R_{p0}(\dot{R}_pt)^2+\frac{1}{4}(\dot{R}_pt)^3]^{1/3}\nonumber \\
&\sim& R_{p0}+\left(\frac{1}{4}\right)^{1/3}\dot{R}_pt
\label{r2}
\end{eqnarray}


\begin{thebibliography}{99}

\bibitem[Armitage(2010)]{Arm 10} Armitage, P.~J.\ 2010, 
Astrophysics of Planet Formation, by Philip J.~Armitage, pp.~294.~ISBN 
978-0-521-88745-8 (hardback).~Cambridge, UK: Cambridge University Press, 
2010., 

\bibitem[Barnes et al.(2009)]{Bar 09} Barnes, R., Quinn, 
T.~R., Lissauer, J.~J., \& Richardson, D.~C.\ 2009, Icarus, 203, 626 

\bibitem[Benz 
\& Asphaug(1999)]{BA 99} Benz, W., \& Asphaug, E.\ 1999, Icarus, 142, 5 

\bibitem[Blum 
\& Wurm(2008)]{BW 08} Blum, J., \& Wurm, G.\ 2008, \araa, 46, 21 

\bibitem[Brauer et 
al.(2008)]{Bra 08} Brauer, F., Henning, T., \& Dullemond, C.~P.\ 2008, \aap, 487, L1

\bibitem[Chambers 
\& Wetherill(1998)]{CW 98} Chambers, J.~E., \& Wetherill, G.~W.\ 1998, Icarus, 136, 304 

\bibitem[Chambers(2004)]{Cha 04} Chambers, J.~E.\ 2004, Earth and Planetary Science Letters, 223, 241

\bibitem[Chambers(2010)]{Cha 10} Chambers, J.~E.\ 2010, Icarus inpress

\bibitem[Chiang 
\& Youdin(2009)]{CY 09} Chiang, E., \& Youdin, A.\ 2009, arXiv:0909.2652 

\bibitem[Cieza et al.(2009)]{Cie 09} Cieza, L.~A., et al.\ 
2009, \apjl, 696, L84

\bibitem[Cuzzi 
\& Weidenschilling(2006)]{CW 06} Cuzzi, J.~N., \& Weidenschilling, S.~J.\ 2006, Meteorites and the Early Solar System II, 353

\bibitem[Cuzzi et al.(2008)]{Cuz 08} Cuzzi, J.~N., Hogan, 
R.~C., \& Shariff, K.\ 2008, \apj, 687, 1432

\bibitem[Cuzzi et al.(2010)]{Cuz 10} Cuzzi, J.~N., Hogan, 
R.~C., \& Bottke, W.~F.\ 2010, arXiv:1004.0270 

\bibitem[Dominik et al.(2007)]{Dom 07} Dominik, C., Blum, J., 
Cuzzi, J.~N., \& Wurm, G.\ 2007, Protostars and Planets V, 783 


\bibitem[Goldreich 
\& Ward(1973)]{GW 73} Goldreich, P., \& Ward, W.~R.\ 1973, \apj, 183, 1051 

\bibitem[Haghighipour(2009)]{Hag 09a} Haghighipour, N.\ 2009, 
arXiv:0908.3328

\bibitem[Haghighipour et al.(2009)]{Hag 09b} Haghighipour, N., 
Dvorak, R., \& Pilat-Lohinger, E.\ 2009, arXiv:0911.0819 

\bibitem[Hayashi(1981)]{Hay 81} Hayashi, C.\ 1981, Progress of 
Theoretical Physics Supplement, 70, 35

\bibitem[Heng 
\& Kenyon(2010)]{HK 10} Heng, K., \& Kenyon, S.~J.\ 2010, arXiv:1005.1660 

\bibitem[Ida 
\& Lin(2004)]{IL 04} Ida, S., \& Lin, D.~N.~C.\ 2004, \apj, 604, 388

\bibitem[Johansen et al.(2007)]{Joh 07} Johansen, A., Oishi, 
J.~S., Low, M.-M.~M., Klahr, H., Henning, T., 
\& Youdin, A.\ 2007, \nat, 448, 1022 

\bibitem[Johansen et 
al.(2008)]{Joh 08} Johansen, A., Brauer, F., Dullemond, C., Klahr, H., \& Henning, T.\ 2008, \aap, 486, 597 

\bibitem[Kenyon 
\& Bromley(2006)]{KB 06} Kenyon, S.~J., \& Bromley, B.~C.\ 2006, \aj, 131, 1837 

\bibitem[Kenyon 
\& Bromley(2010)]{KB 10} Kenyon, S.~J., \& Bromley, B.~C.\ 2010, \apjs, 188, 242

\bibitem[Kokubo 
\& Ida(1996)]{KI 96} Kokubo, E., \& Ida, S.\ 1996, Icarus, 123, 180 

\bibitem[Kokubo 
\& Ida(1998)]{KI 98} Kokubo, E., \& Ida, S.\ 1998, Icarus, 131, 171 

\bibitem[Kokubo et al.(2006)]{Kok 06} Kokubo, E., Kominami, 
J., \& Ida, S.\ 2006, \apj, 642, 1131

\bibitem[Larwood et al.(1996)]{Lar 96} Larwood, J.~D., Nelson, 
R.~P., Papaloizou, J.~C.~B., \& Terquem, C.\ 1996, \mnras, 282, 597

\bibitem [Lissauer(1993)]{Lis 93} Lissauer, J.~J.\ 1993, \araa, 31, 129 

\bibitem[Leinhardt 
\& Richardson(2005)]{LR 05} Leinhardt, Z.~M., \& Richardson, D.~C.\ 2005, \apj, 625, 427 

\bibitem[Levison 
\& Agnor(2003)]{LA 03} Levison, H.~F., \& Agnor, C.\ 2003, \aj, 125, 2692 

\bibitem[Marzari et 
al.(2009)]{Mar 09} Marzari, F., Th{\'e}bault, P., \& Scholl, H.\ 2009, \aap, 507, 505 

\bibitem[Morbidelli et al.(2009)]{Mor 09} Morbidelli, A., 
Bottke, W.~F., Nesvorn{\'y}, D., \& Levison, H.~F.\ 2009, Icarus, 204, 558 

\bibitem[Natta et al.(2007)]{Nat 07} Natta, A., Testi, L., 
Calvet, N., Henning, T., Waters, R., 
\& Wilner, D.\ 2007, Protostars and Planets V, 767

\bibitem[Paardekooper et al.(2008)]{Par 08} Paardekooper, 
S.-J., Th{\'e}bault, P., \& Mellema, G.\ 2008, \mnras, 386, 973 

\bibitem[Paardekooper 
\& Leinhardt(2010)]{PL 10} Paardekooper, S.~-., \& Leinhardt, Z.~M.\ 2010, arXiv:1001.3025 

\bibitem[Pourbaix et 
al.(2002)]{Pou 02} Pourbaix, D., et al.\ 2002, \aap, 386, 280

\bibitem[Safronov(1969)]{Saf 69}
Safronov, V. S. 1969. Evolution of the Protoplanetary
Cloud and Formation of the Earth and Planets. Moscow: Nauka. Eng.
trans. NASA TTF-677, 1 972

\bibitem[Scott(2007)]{Sco 07} Scott, E.~R.~D.\ 2007, Annual 
Review of Earth and Planetary Sciences, 35, 577

\bibitem[Teiser 
\& Wurm(2009)]{TW 09} Teiser, J., \& Wurm, G.\ 2009, \mnras, 393, 1584 

\bibitem[Th{\'e}bault et al.(2006)]{The 06} Th{\'e}bault, P., 
Marzari, F., \& Scholl, H.\ 2006, Icarus, 183, 193 

\bibitem[Th{\'e}bault et al.(2008)]{The 08} Th{\'e}bault, P., 
Marzari, F., \& Scholl, H.\ 2008, \mnras, 388, 1528

\bibitem[Th{\'e}bault et al.(2009)]{The 09} Th{\'e}bault, P., 
Marzari, F., \& Scholl, H.\ 2009, \mnras, 393, L21 

\bibitem[Wettlaufer(2009)]{Wet 09} Wettlaufer, J.~S.\ 2009, 
arXiv:0911.5398 

\bibitem[Weidenschilling(1977)]{Wei 77} Weidenschilling, 
S.~J.\ 1977, \mnras, 180, 57 

\bibitem[Weidenschilling(1980)]{Wei 80} Weidenschilling, 
S.~J.\ 1980, Icarus, 44, 172

\bibitem[Weidenschilling 
\& Cuzzi(1993)]{WC 93} Weidenschilling, S.~J., \& Cuzzi, J.~N.\ 1993, Protostars and Planets III, 1031 

\bibitem[Weidenschilling(1997)]{Wei 97} Weidenschilling, 
S.~J.\ 1997, From Stardust to Planetesimals, 122, 281

\bibitem[Weidenschilling et al.(1997)]{Wei 97b} 
Weidenschilling, S.~J., Spaute, D., Davis, D.~R., Marzari, F., 
\& Ohtsuki, K.\ 1997, Icarus, 128, 429 

\bibitem[Weidenschilling(2010)]{Wei 10} Weidenschilling, 
S.~J.\ 2010, Lunar and Planetary Institute Science Conference Abstracts, 
41, 1453

\bibitem[Wetherill 
\& Inaba(2000)]{WI 00} Wetherill, G.~W., \& Inaba, S.\ 2000, Space Science Reviews, 92, 311 

\bibitem[Xie 
\& Zhou(2008)]{XZ 08} Xie, J.-W., \& Zhou, J.-L.\ 2008, \apj, 686, 570 

\bibitem[Xie 
\& Zhou(2009)]{XZ 09} Xie, J.-W., \& Zhou, J.-L.\ 2009, \apj, 698, 2066 

\bibitem[Xie et al.(2010)]{Xie 10} Xie, J.-W., Zhou, J.-L., 
\& Ge, J.\ 2010, \apj, 708, 1566

\bibitem[Youdin 
\& Goodman(2005)]{YG 05} Youdin, A.~N., \& Goodman, J.\ 2005, \apj, 620, 459

\bibitem[Youdin 
\& Shu(2002)]{YS 02} Youdin, A.~N., \& Shu, F.~H.\ 2002, \apj, 580, 494 

\bibitem[Youdin(2008)]{You 08} Youdin, A.\ 2008, 
arXiv:0807.1114

\end{thebibliography}
\end{document}